# Hypothesis: Homologous recombination depends on parallel search

*Johan Elf, Dept. Cell and Molecular Biology, Uppsala University*

Having a double stranded break (DSB) of chromosomal DNA is a serious condition for any cell. In all kingdoms of life cells achieve error free chromosome repair after DSB by using a reference sequence in a sister chromatid or a homologous chromosome through the process of homologous recombination (HR). The molecular mechanism of HR has been worked out in great detail [1], since HR-deficiency is a major source of cancer [2]. However, the mystery of how a broken sequence can find its homologous reference sufficiently fast is unsolved [3].

The problem is that the homologous reference sequence can be anywhere in the cell [4]. Thus, HR has to depend on probing each conceivable chromosome position although this requires unwinding of the double helix and base pairing with the broken chromosome sequence. Even considering recent in vitro measurements revealing rapid sampling with short oligos [5] it would take more than a year to unwind and compare all sequences in a mammalian genome and weeks for yeast genome. If we add the diffusive and topological constraints of moving and aligning the broken chromosome in a dense nucleus, the time scales involved become astronomical, especially considering that the broken ends must also be held together[6]. Still the cell manages to find the homologous sequence in hours [7]. Something fundamental is clearly missing in our understanding of homologous recombination.

If it takes too long to solve a problem sequentially it can often be parallelized. Thus, I suggest that the cell makes many short copies of the sequences flanking the DBS and use these to search for the homologous sequence in parallel. This strategy not only parallelizes the search process but also eliminates the need to move the broken chromosome around in a time consuming manner. Although the necessary molecular components to implement such a strategy have already have been characterized, the transformative impact of this strategy on search speed has not been discussed.

For example, in Arabidopsis and in human cells, 21 nt RNAs are made, corresponding to the sequences flanking the DBS [8]. These so called diRNA bind to Ago2, which from separate studies, e.g. ref[9], are seen targeting chromosomal promoters or nascent transcripts when programmed by homologous 21 nt microRNA. Thus, the Ago2-diRNA complexes are ideal candidates for searching the genome for sequences homologous to the broken sequence. Ago2 deletions are further known to be deficient in HR and Ago2 is directly binding to the critical HR protein Rad51 even in the absence of the diRNA[10].

An RNA corresponding to the diRNA has not been described in bacteria, but the many short ssDNA resection products that are made by the recBCD or addAB systems while processing the DSB ends [11], could play the same role. This would also explain why ssDNA degradation products have lengths of tens of nucleotides[12]. In fact, short pieces of ssDNA bind RecA in vitro and the complex also search for specific sequences in dsDNA[13], but it is not known whether the resection products of recBCD and addAB are bound by RecA in vivo. If they are, they would recruit RecA to the region homologous to the break site.

Once the homologous sequence has been located, there are many ways to make the region more accessible for the recombination with the broken chromosome. This second step may e.g. involve formation of RecA / Rad51 polymers [14], possibly nucleated at the homologous sequences, or alternatively relocation of any RecA / Rad51 bound DNA to nuclear "repair centers" [15]. In any case, given that the homologous region already is found and marked through parallel search by many freely diffusing molecules, the overall process can be made many thousands times faster.

In summary, I propose that the solution to the homologous recombination search problem is a parallelized search implemented by freely diffusing molecules programmed with sequences corresponding to those flanking the break site.

## Acknowledgment

I am grateful to Ran Kafri for introducing me to the existence diRNA and to Claudia Kutter for helpful discussion.